\documentstyle[osa,12pt]{revtex}
\begin{document}
\title{ANTIQUARK POLARIZATION INSIDE THE PROTON IS SMALL\thanks{%
\ \ CMU-HEP95-15, DOE-ER/40682-104, hep-ph/9510\#\#\#}}
\author{{\normalsize \ }{\large \ T. P. Cheng$^{\dagger }$ and Ling-Fong Li}$%
^{\diamond }$}
\address{$^{\diamond }$Department of Physics, Carnegie Mellon University,
Pittsburgh,%
\\
PA 15213\\
{\em \ }$^{\dagger }$Department of Physics and Astronomy, University of\\
Missouri, St Louis, MO 63121}
\maketitle

\begin{abstract}
Quark contributions to the proton spin as deduced from polarized DIS of
leptons off a nucleon target, and the octet baryon magnetic moments, can be
used to deduce the antiquark polarizations $\Delta _{\overline{q}}$ inside
the proton. In this way the 1992 analysis by Karl is shown to imply $\Delta
_{\overline{q}}\simeq 0.$ Such a spin structure fits nicely into the chiral
quark interpretation of the proton spin and flavor puzzles.
\end{abstract}

\ \

\section{Introduction}

In the last several years, starting with EMC\cite{emc} in the late 1980's,
the polarized deep inelastic lepton-nucleon scattering experiments have
revealed a nucleon spin structure that deviates significantly from the
simple quark model expectation. It is particularly puzzling in view of the
fact that the same spin structure in the nonrelativistic constituent quark
model seems to lead to a reasonably good description of baryon magnetic
moments\cite{perkins}. In this paper we re-examine the proton spin data in
connection with the octet baryon magnetic moments. We find that such an
investigation can lead to some useful insight into the proton spin
structure, above and beyond what the (spin-dependent structure function) $g_1
$ sum rule\cite{bjsr, ejsr} and the baryonic weak axial vector couplings can
tell us. Namely, this combined analysis suggests that the antiquarks inside
the proton are not significantly polarized. And, as we shall show, this
lends further support to the chiral quark model\cite{mgtheor} approach to
the various nucleon structure puzzles\cite{ehq, clprl}.

In Sec.2, we first discuss the simple model where the magnetic moment of a
baryon comes from those of its constituent quarks. The flavor-$SU(3)$\ is
used to express the octet baryon moments in terms of the intrinsic $u,\;d,\;$%
and $s$ magnetic moments and the quark and antiquark contributions to the
proton spin. In Sec.3, the ''proton spin crisis'' is briefly recalled. We
then show, in Sec.4, that an analysis made by Karl\cite{gk} actually
contains definite information about the antiquark polarizations in the
proton. In Sec.5 the spin and magnetic moment structure is studied in the
chiral quark model. Finally we conclude and discuss the implication of our
result.

\section{Spin structure and magnetic moments}

Let us first discuss the relation between the quark and antiquark
polarization contributions to the proton spin and the proton magnetic
moment. One usually denotes the $q$-flavor contribution to the proton spin
as:
\begin{equation}
\Delta q=\left( q_{\uparrow }-q_{\downarrow }\right) +\left( \overline{q}%
_{\uparrow }-\overline{q}_{\downarrow }\right) \equiv \Delta _q+\Delta _{%
\overline{q}}  \label{delsum}
\end{equation}
where $q_{\uparrow }$ is the number of $q$-flavor quarks with spin parallel,
and $q_{\downarrow }$ anti-parallel, to the proton spin. Thus $\Delta _q$
and $\Delta _{\overline{q}}$ are the quark and antiquark polarizations,
respectively.

For the $q$-flavor quark contribution to the proton magnetic moment, we have
however
\begin{equation}
\mu _p\left( q\right) =\Delta _q\mu _q+\Delta _{\overline{q}}\mu _{\overline{%
q}}=\left( \Delta _q-\Delta _{\overline{q}}\right) \mu _q\equiv \widetilde{%
\Delta q}\mu _q  \label{deldiff}
\end{equation}
where $\mu _q$ is the magnetic moment of the $q$-flavor quark. The negative
sign simply reflects the opposite quark and antiquark moments, $\mu _{%
\overline{q}}=-\mu _q$, as in general $\mu _q=e_qg_q/2m_q$, and when we go
from quark to antiquark only the charge $e_q\;$changes sign, but not the
mass $m_q$ or the gyromagnetic ratio $g_q.$ Thus the spin factor that enters
into the expression for the magnetic moment is $\widetilde{\Delta q}$, the
{\em difference} of the quark and antiquark polarizations. If we assume that
the proton magnetic moment is entirely built up from the light quarks inside
it, we have
\begin{equation}
\mu _p=\widetilde{\Delta u}\mu _u+\widetilde{\Delta d}\mu _d+\widetilde{%
\Delta s}\mu _s.  \label{proton}
\end{equation}
In such an expression there is a {\em separation} of the intrinsic quark
magnetic moments and the spin wavefunctions. Flavor-$SU(3)$ symmetry then
implies, the proton wavefunction being related the $\Sigma ^{+}$
wavefunction by the interchange of $d\leftrightarrow s$ and $\overline{d}%
\leftrightarrow \overline{s}\;$quarks, the relations $\left( \widetilde{%
\Delta u}\right) _{\Sigma ^{+}}=\left( \widetilde{\Delta u}\right) _p\equiv
\widetilde{\Delta u},$ $\left( \widetilde{\Delta d}\right) _{\Sigma ^{+}}=%
\widetilde{\Delta s},$ and$\;\left( \widetilde{\Delta s}\right) _{\Sigma
^{+}}=\widetilde{\Delta d};$ similarly it being related to the $\Xi ^0$
wavefunction by a further interchange of $u\leftrightarrow s$ quarks, thus $%
\left( \widetilde{\Delta d}\right) _{\Xi ^0}=\left( \widetilde{\Delta d}%
\right) _{\Sigma ^{+}}=\widetilde{\Delta s},\;\left( \widetilde{\Delta s}%
\right) _{\Xi ^0}=\left( \widetilde{\Delta u}\right) _{\Sigma ^{+}}=%
\widetilde{\Delta u},$ and $\left( \widetilde{\Delta u}\right) _{\Xi
^0}=\left( \widetilde{\Delta s}\right) _{\Sigma ^{+}}=\widetilde{\Delta d}.$
We have,
\begin{eqnarray}
\mu _{\Sigma ^{+}} &=&\widetilde{\Delta u}\mu _u+\widetilde{\Delta s}\mu _d+%
\widetilde{\Delta d}\mu _s,  \label{sigma+} \\
\mu _{\Xi ^0} &=&\widetilde{\Delta d}\mu _u+\widetilde{\Delta s}\mu _d+%
\widetilde{\Delta u}\mu _s,  \label{cascade0}
\end{eqnarray}
the intrinsic moments $\mu _q^{\prime }s$ being unchanged when we go from
Eq.(\ref{proton}) to Eqs.(\ref{sigma+}) and (\ref{cascade0}). The $%
n,\;\Sigma ^{-}$, and $\Xi ^{-}$ moments can be obtained from their isospin
conjugate partners $p,\Sigma ^{+}$, and $\Xi ^0$ by the interchange of their
respective $u\leftrightarrow d$ quarks: $\left( \widetilde{\Delta u}\right)
_{\Sigma ^{-}}=\left( \widetilde{\Delta d}\right) _{\Sigma ^{+}}=\widetilde{%
\Delta s},\;etc.$%
\begin{eqnarray}
\mu _n &=&\widetilde{\Delta d}\mu _u+\widetilde{\Delta u}\mu _d+\widetilde{%
\Delta s}\mu _s,  \label{neutron} \\
\mu _{\Sigma ^{-}} &=&\widetilde{\Delta s}\mu _u+\widetilde{\Delta u}\mu _d+%
\widetilde{\Delta d}\mu _s,  \label{sigma-} \\
\mu _{\Xi ^{-}} &=&\widetilde{\Delta s}\mu _u+\widetilde{\Delta d}\mu _d+%
\widetilde{\Delta u}\mu _s.  \label{cascade-}
\end{eqnarray}
The relations for the $I_z=0,\;Y=0$ moments are more complicated in
appearance but the underlying arguments are the same.
\begin{equation}
\mu _\Lambda =\frac 16\left( \widetilde{\Delta u}+4\widetilde{\Delta d}+%
\widetilde{\Delta s}\right) \left( \mu _u+\mu _d\right) +\frac 16\left( 4%
\widetilde{\Delta u}-2\widetilde{\Delta d}+4\widetilde{\Delta s}\right) \mu
_s,  \label{lamda}
\end{equation}

\begin{equation}
\mu _{\Lambda \Sigma }=\frac{-1}{2\sqrt{3}}\left( \widetilde{\Delta u}-2%
\widetilde{\Delta d}+\widetilde{\Delta s}\right) \left( \mu _u-\mu _d\right)
.  \label{transmom}
\end{equation}
The proton Eq.(\ref{proton}) and neutron Eq.(\ref{neutron}) were first
written down by Sehgal\cite{sehgal}, and its generalization to other octet
baryons have been discussed by a number of authors\cite{br}. Since we will
be following most closely Karl's work\cite{gk}, we shall refer to Eqs.(\ref
{proton}) to (\ref{transmom}) as the Karl-Sehgal (KS) equations. We should
however note that the equations actually written down by Karl involve the
polarization sum $\Delta q=\Delta _q+\Delta _{\overline{q}}$ rather than the
difference $\widetilde{\Delta q}=\Delta _q-\Delta _{\overline{q}}$ because
he has chosen to work with an ''effective quark magnetic moment'' $\mu
_q^{\prime }=\widetilde{\Delta q}\mu _q/\Delta q.$ The resultant equation $%
\mu _p(q)=\Delta q\mu _q^{\prime },$ instead of Eq.(\ref{deldiff}), thus
does not separate the spin-wavefunction and the quark intrinsic moments.
Namely, the $\mu _q^{\prime }$ moments, just as $\Delta q$ and $\widetilde{%
\Delta q},$ depend on the spin wavefunction as well as the intrinsic
moments. Thus different approaches have been adopted in Ref. 8 and the
present paper. We are of the opinion that, in the effort to express all the
baryon moments in terms of the proton spin factors, our assumption of a
complete separation of the spin wavefunctions and the intrinsic quark
moments is the more reasonable approach.

\section{Simple quark model results and the measured proton spin structure}

In the nonrelativistic constituent quark model, there is no quark sea. The
proton spin follows simply from the addition of its valence quark spins. One
finds:
\begin{equation}
\Delta u=\frac 43,\;\;\;\Delta d=-\frac 13,\;\;\;\Delta s=0,\;\;\;\Delta
\Sigma =1,  \label{qmspin}
\end{equation}
where $\Delta \Sigma =\Delta u+\Delta d+\Delta s$ is the total quark
contribution to the proton spin. There is no strange quark inside proton,
hence no strange quark polarization, $\Delta s=0$. Also, there is no
antiquark, hence no antiquark polarization, $\Delta _{\overline{q}}=0.$ This
means that in the simple quark model (sQM) we have $\Delta q=\widetilde{%
\Delta q}.\;$The proton and neutron magnetic moments, for example, are given
by Eqs.(\ref{proton}), (\ref{neutron}), and (\ref{qmspin}), as
\begin{eqnarray}
\mu _p &=&\frac 43\mu _u-\frac 13\mu _d=\frac{m_N}{m_{u,d}}g\mu _N,
\label{qmpro} \\
\mu _n &=&-\frac 13\mu _u+\frac 43\mu _d=-\frac 23\frac{m_N}{m_{u,d}}g\mu _N,
\label{qmneut}
\end{eqnarray}
where we have assumed the equality of $u$ and $d$ constituent quark masses,
denoted by $m_{u,d}$, as well as their gyromagnetic ratios $g_u=g_d=g$ (for
example, $g=2$ for Dirac particles). $\mu _N$ is the nucleon magneton, and $%
m_N$ the nucleon mass. From this we deduce the famous $SU(6)$ result\cite
{blpais} for the moment ratio of $\mu _p/\mu _n=-1.5$, which is to be
compared to the experimental value of $\left( \mu _p/\mu _n\right) _{\text{%
exptl}}=-1.48.\;$Furthermore with a measured value of $\mu _p=\ 2.79\mu _N,$
Eq.(\ref{qmpro}) suggests that $m_{u,d}$ is about a third of the nucleon
mass. In this way the simple constituent quark model actually provides a
reasonable fit to all the octet moments (Table 1), especially if we allow
for a heavier strange constituent quark mass, $m_{u,d}/m_s\simeq 0.6.$

However, in 1988, EMC reported\cite{emc} their measurement of the proton
structure function $g_1(x),$ which seemed to suggest a quark contribution to
the proton spin very different from this simple quark model expectation (\ref
{qmspin}). Using $SU(3)$ symmetry and baryonic weak axial-vector couplings $%
g_A=1.254$, $F/D=0.632$ and the $g_1$ sum rule, one could deduce the various
quark contributions to the proton spin:
\begin{equation}
\Delta u=0.75\pm 0.12,\;\;\;\Delta d=-0.51\pm 0.12,\;\;\;\Delta s=-0.22\pm
0.12,\;\;\;\Delta \Sigma =0.02\pm 0.21.\;  \label{emcspin}
\end{equation}
$\;$

The discrepancy between the phenomenological values (\ref{emcspin}) and the
quark model expectations (\ref{qmspin}) --- especially the indication of the
presence in the proton a strongly polarized strange quark sea, and the
possibility that proton gets almost none of its spin from quarks --- caused
it to be called the ''proton spin crisis''. It is certainly puzzling why the
simple quark model spin structure (\ref{qmspin}) can lead to a satisfactory
description of the baryon magnetic moments, and yet fails to be in agreement
with a more direct measurement (\ref{emcspin})?

\section{A combined analysis of the proton spin and baryon magnetic moment
data}

Instead of viewing this discrepancy as a definitive failure of the
constituent quark model, we consider it as showing that for certain
phenomena the effects of the quark sea can be very important. The issue is
then what are the properties of the quark sea do the observations imply, and
what mechanism can produce such a quark sea. We may start with the following
questions:

(i) Does a good fit to the baryon magnetic moments necessarily mean that the
quark contributions to the proton spin must have the sQM values (\ref{qmspin}%
)? Or, does the measured spin factors (\ref{emcspin}) yield a good (or even
better) fit to the magnetic moments?

(ii) As both spin factors $\Delta q^{\prime }s$ and magnetic moments $\mu
_B^{\prime }s$ are related to the quark polarizations inside the proton, can
an analysis using $\left( \Delta q\right) _{\text{exptl}}$ {\em and} $\mu _B$
values lead to further insights into the proton spin structure, above and
beyond what each set can reveal?

It turned out that the analysis performed in l992 by Karl\cite{gk} had gone
a long way in answering these questions. What Karl did was to use the KS
equations: (\ref{proton}) to (\ref{transmom}), to search for the values of $%
\left( \widetilde{\Delta q}\right) ^{\prime }s$ so that the best-fit (in the
sense of lowest $\chi ^2$, etc.) to all the measured baryon magnetic moments
could be obtained. He found the following set:
\begin{equation}
\widetilde{\Delta u}=0.86\pm 0.12,\;\;\;\widetilde{\Delta d}=-0.40\pm
0.12,\;\;\;\widetilde{\Delta s}=-0.20\pm 0.12,\;\;\;\widetilde{\Delta \Sigma
}=0.27\pm 0.21.  \label{karlspin}
\end{equation}
The fit it produces is {\em better} than the simple quark model fit. See
Table 1 for more detail. Thus the answer to Question-(i) is that the
required magnetic moment best-fit values (\ref{karlspin}) are actually
closer to the measured spin values (\ref{emcspin}) than $\left( \Delta
q\right) _{\text{sQM}}$ of (\ref{qmspin}).

In the meantime, the EMC result has been extended by further analysis\cite
{emc}, by new measurements by SMC, E142 and E143\cite{nuspin}, extending to
neutron target, to larger kinematic regime, with better statistics. With
even higher order QCD corrections\cite{qcdc} included, we now know that
Bjorken sum rule\cite{bjsr} is verified\cite{ellisk} to an accuracy about $%
12\%,$ and, although the original EMC result (\ref{emcspin}) has been
confirmed in general terms, with improved accuracy the values of the proton
spin components have been slightly modified. In particular, the total quark
contribution to the proton spin is no longer being consistent with zero. The
more recent result\cite{ellisk} is
\begin{equation}
\Delta u=0.83\pm 0.06,\;\;\;\Delta d=-0.42\pm 0.06,\;\;\;\Delta s=-0.10\pm
0.06,\;\;\;\Delta \Sigma =0.31\pm 0.11  \label{newspin}
\end{equation}
which is even more similar-in-value to the $\widetilde{\Delta q}^{\prime }s$
of Eq.(\ref{karlspin}) than the values known in 1992 when Karl performed his
best-fit analysis. This closeness-in-value between Eq.(\ref{newspin})'s $%
\left( \Delta q\right) _{\text{exptl}}$ and Eq.(\ref{karlspin})'s $\left(
\widetilde{\Delta q}\right) _{\mu _B}$ immediately allows us to infer that
antiquark polarization inside the proton $\Delta _{\overline{q}}=\left(
\overline{q}_{\uparrow }-\overline{q}_{\downarrow }\right) $ is small. In
more quantitative terms, the antiquark to quark polarization fraction $%
\delta _{\overline{q}}$ can be expressed as the difference and sum ratio of
the two kinds of spin factors, $\Delta q\;$and$\;\widetilde{\Delta q}.\;$%
{}From Eqs.(\ref{delsum}) and (\ref{deldiff}), we have
\begin{equation}
\delta _{\overline{q}}=\frac{\Delta _{\overline{q}}}{\Delta _q}=\left| \frac{%
\Delta q-\widetilde{\Delta q}}{\Delta q+\widetilde{\Delta q}}\right| .
\label{aqpol}
\end{equation}
After substituting in the values (\ref{karlspin}) and (\ref{newspin}), and
combing errors by quadrature, we have
\begin{equation}
\delta _{\overline{u}}=0.02\pm 0.08,\;\;\;\delta _{\overline{d}}=0.02\pm
0.16,\;\;\;\delta _{\overline{s}}=0.33\pm 0.45.  \label{aqpols}
\end{equation}
They are all consistent with being zero, although the error on the
anti-strange quark polarization is quite large. Still, this result is
suggestive that the polarization of the antiquarks in the quark-sea is
suppressed.

As explained at the end of Sec.2, Eqs.(\ref{proton}) to (\ref{transmom}) as
written down by Karl\cite{gk} have $\Delta q^{\prime }s$ in place of $%
\widetilde{\Delta q}^{\prime }s.$ Thus he was not able to make the
connection to the antiquark polarization result as discussed in this paper.

\section{Spin and magnetic moment in the chiral quark model}

The measured values $\Delta q^{\prime }s$ in Eq.(\ref{newspin}) are all
smaller than the sQM prediction (\ref{qmspin}). Thus the quark sea must have
the following specific features: for each flavor, the quark sea must be
strongly polarized in the direction opposite to the proton spin, and yet the
antiquarks in this sea are not much polarized. We shall show that the chiral
quark model produces just such a quark sea.

The basic idea of chiral quark model is that the energy scale associated
with chiral symmetry breaking $\Lambda _{\chi SB}\simeq 1\;GeV$ is much
larger than that associated with QCD confinement $\Lambda _{QCD}\simeq 0.1$-$%
0.3\;GeV.$ The distance scale in between these two nonperturbative QCD
thresholds just corresponds to the interior of a hadron (but not so short a
distance when perturbative QCD becomes operative). In this intermediate
non-perturbative QCD regime, the relevant degrees of freedom are the {\em %
quasiparticles} of quarks, gluons, and the Goldstone bosons associated with
the spontaneous breaking of the $SU(3)\times SU(3)$ chiral symmetry. Here,
the quarks propagate in a ground state filled with the collective
excitations of $q\overline{q}$ condensates and in this way gain a large
constituent quark mass. The Goldstone bosons are the usual pseudoscalar
mesons, but propagating here in the interior of the hadron. (we shall refer
to them as the {\em internal }Goldstone bosons.) The quark-gluon
interactions of the underlying QCD bring about chiral symmetry breaking and
the Goldstone modes of excitations. However, when the description is
reorganized in terms of the quasiparticle effective fields of constituent
quarks and Goldstone bosons, we expect that the gluons will now have a
negligibly small effective coupling. Thus, the only important interaction is
the coupling among the Goldstone bosons and quarks.

A quark sea created through internal Goldstone boson (GB) emissions by a
valence quark,
\begin{equation}
q_{\uparrow }\rightarrow GB+q_{\downarrow }^{\prime }\rightarrow \left( q\;%
\overline{q^{\prime }}\right) _0\;q_{\downarrow }^{\prime }  \label{gbemit}
\end{equation}
has just the desired spin polarization features. The coupling of the
pseudoscalar Goldstone boson to the quarks will flip the polarization of the
quark: $q_{\uparrow }$ $\rightarrow $ $q_{\downarrow }^{\prime }.$ We note
that the final state $q_{\downarrow }^{\prime }$ carries {\em all} the
polarization of the produced quark-sea, as the pair $\left( q\;\overline{%
q^{\prime }}\right) _0$ --- coming out of the Goldstone boson --- must be in
the spin-zero combination:
\begin{equation}
\left( q\;\overline{q^{\prime }}\right) _0=\frac 1{\sqrt{2}}\left(
q_{\uparrow }\overline{q_{\downarrow }^{\prime }}-q_{\downarrow }\overline{%
q_{\uparrow }^{\prime }}\right) .  \label{zerospin}
\end{equation}
The quark sea created by such a mechanism will be polarized, as given by $%
q_{\downarrow }^{\prime },$ in a direction opposite to the proton spin, and
the produced antiquark $\overline{q^{\prime }}$ must be unpolarized ({\em %
i.e.} equal probability for $\overline{q_{\uparrow }^{\prime }}$ and $%
\overline{q_{\downarrow }^{\prime }}$) as shown in (\ref{zerospin}).

In a previous publication \cite{clprl} we have shown that the broken-$U(3)$
version of the chiral quark model with two parameters (the octet and singlet
Goldstone bosons couplings to the quarks) can provide a simple and unified
account of the proton's spin and flavor puzzles. The quark contributions to
the proton spin have been calculated to be
\begin{equation}
\Delta u=\frac 43-\frac 19(37+8\varsigma ^2)a,\;\;\Delta d=-\frac 13-\frac 29%
(1-\varsigma ^2)a,\;\;\Delta s=-a.  \label{cqmspin}
\end{equation}
$\;\;$where $a\propto \left| g_8\right| ^2$ is the probability for a $u$%
-quark to emit a $\pi ^{+}$ [and its $SU(3)$ generalizations], $\varsigma
=g_1/g_8$ is the singlet and octet Goldstone boson coupling ratio. We have
shown that with a choice, for example, of $a=0.1$ and $\varsigma =-1.2$,
this model yields a $\overline{u}$-$\overline{d}$ asymmetry compatible with
the observed violation of the Gottfried sum rule\cite{nmc}, and the observed
asymmetry in the proton-neutron Drell-Yan processes\cite{na51}, {\em etc. }%
Such parameters then yield, through Eq.(\ref{cqmspin}) the spin values as
\begin{equation}
\Delta u=0.79,\;\;\,\;\Delta d=-0.32,\;\;\,\;\Delta s=-0.10,\;\;\;\,\Delta
\Sigma =0.37,  \label{cqmdels}
\end{equation}
to be compared to the phenomenological values (\ref{newspin}).

Before discussing all the magnetic moments, let us first show that the
well-known $SU(6)$ result for the proton-neutron ratio is maintained in the
chiral quark model. Substituting (\ref{cqmspin}) into Eqs.(\ref{proton}) and
(\ref{neutron}) we have
\begin{equation}
\frac{\mu _p}{\mu _n}=\left( -\frac 32\right) \left[ 1-\frac 56a\left( 1-%
\frac{m_{u,d}}{m_s}\right) \right]  \label{cqmpnr}
\end{equation}
{\em i.e. }the $SU(6)$ result is preserved in the flavor-$SU(3)$ limit of $%
m_{u,d}=m_s.$

That our result (\ref{cqmdels}) is reasonably close to the $\left( \Delta
q\right) _{\text{exptl}}^{\prime }$s of (\ref{newspin}), which are in turn
close-in-value to $\left( \widetilde{\Delta q}\right) _{\mu _B}^{\prime }$s
of (\ref{karlspin}), leads us to expect that this model should be able to
give a satisfactory description of the baryon magnetic moments as well. In
the last column of Table 1 the numerical values for the illustrative
parameters of $a=0.10$ and $\varsigma =-1.2$ are presented. One should keep
in mind that our's is an $SU(3)$ symmetric calculation, as it is based on KS
equations and the $SU(3)\times U(1)$ symmetric chiral quark model. The only $%
SU(3)$ breaking effect that has been taken into account is the different
constituent masses $m_{u,d}\neq m_s$ in the quark intrinsic moments. Thus we
do not really expect a better than 20 to 30\% agreements from the model
predictions.

\section{Discussion}

The KS equations, whether used in the simple constituent quark model, the
chiral quark model, or the best-fit program by Karl, give a reasonably good
account of $\mu _B^{\prime }s.$ We interpret this fact to mean that the
basic idea of the baryon magnetic moment being built up from the constituent
quark moments is a robust framework. We have thus neglected possible
contribution to the baryon magnetic moment by the orbital motion of the
quarks. The success of the fits may be viewed as an {\em a posteriori}
justification of this assumption. In this framework, we have shown that the
antiquarks inside the proton are not significantly polarized. And, such a
spin structure fits nicely into the chiral quark model approach where the
antiquarks are produced through the emissions of the spin-zero internal
Goldstone bosons.

Many authors\cite{gluon} have suggested that the deviation of the
''observed'' quark spin contribution $\left( \Delta q\right) _{\text{exptl}}$
from the simple quark model prediction $\left( \Delta q\right) _{\text{sQM}}$
is due to a possible gluonic polarization $\Delta G.\;$These authors argued
that one must first subtract out this gluonic term in order to reveal the
''true'' quark contributions, $\left( \Delta q\right) _{\text{exptl}}=\left(
\Delta q\right) _{\text{true}}-\frac{\alpha _s}{2\pi }\Delta G$, and they
suggest that only then do we expect agreement with the quark model
expectation, $\left( \Delta q\right) _{\text{true}}=\left( \Delta q\right) _{%
\text{sQM}}.\;$According to this approach, one would then predict, because
of the simple quark model property of $\Delta q=\widetilde{\Delta q},\;$that
the best-fit $\left( \widetilde{\Delta q}\right) _{\mu _B}^{\prime }$s
should be close in value to the $\left( \Delta q\right) _{\text{sQM}%
}^{\prime }$s of (\ref{qmspin}). Karl has shown that (\ref{qmspin}) actually
is not as a good fit as the values of in (\ref{karlspin}). We interpret it
to mean that this particular scenario with a significant $\Delta G\;$is not
favored by the magnetic moment data.

On the other hand, the broken-$U(3)$ version of the chiral quark model\cite
{clprl} can account for all these spin and magnetic moment data, as well as
the flavor structure, in a simple and natural way. This list of positive
attributes of the chiral quark model has been enhanced by the independent
suggestion that Goldstone boson exchanges can provide a better description
of the fine-structure features of the hadron spectroscopy\cite{finland}. To
us, all these results suggest that the original nonrelativistic quark model
is basicly correct in its description of the low energy hadron physics. It
only needs to be augmented by a light quark sea which is generated
perturbatively by the valence quarks through internal Goldstone boson
emissions. This approach to the fundamental problem of proton structure
warrants further investigation.

This work is supported at CMU by the US Department of Energy
(DE-FG02-91ER-40682), and at UM-St Louis by the National Science Foundation
(PHY-9207026).

\begin{center}
\
\end{center}

\begin{center}
\newpage\ \

TABLE 1\ Different fits to the baryon magnetic moments {\em via} KS
equations.
\end{center}

The simple quark model and Karl's best-fit results are from Ref.\cite{gk},
with the quark model having a $\chi ^2/DF=7.35/5$ compared to the best-fit
with $4.42/4.$ Here the $\mu _q^{\prime }s$ are among the parameters being
varied to produce the fits. Thus no specific assumption about the
gyromagnetic ratio such as $g_q=2$ has been made. The last column is meant
to illustrate the ability of the chiral quark model to account for $\mu
_B^{\prime }$s. We have taken $\Delta q=\widetilde{\Delta q}$ with values as
given in (\ref{cqmdels}), and have, with the constraint of $\mu _u=-2\mu _d$
and$\,\mu _s/\mu _d=0.6$, adjusted the remaining one independent value of
the $\mu _q^{\prime }s$ to get a good fit.

\begin{center}
\smallskip\

\begin{tabular}{|ccccc|}
\hline\cline{5-5}\cline{1-1}\cline{2-4}
\multicolumn{1}{|c|}{} & magnetic & simple & \multicolumn{1}{c|}{Karl's} &
chiral quark model \\
\multicolumn{1}{|c|}{} & moments & quark model & \multicolumn{1}{c|}{best-fit
} & $a=0.1,\;\varsigma =-1.2$ \\ \hline\hline
\multicolumn{1}{|c|}{$p$} & $2.79$ & $2.68$ & \multicolumn{1}{c|}{$2.69$} & $%
2.69$ \\
\multicolumn{1}{|c|}{$n$} & $-1.91$ & $-1.92$ & \multicolumn{1}{c|}{$-1.85$}
& $-1.88$ \\
\multicolumn{1}{|c|}{$\Sigma ^{+}$} & $2.48$ & $2.55$ & \multicolumn{1}{c|}{$%
2.59$} & $2.56$ \\
\multicolumn{1}{|c|}{$\Sigma ^{-}$} & $-1.16$ & $-1.13$ &
\multicolumn{1}{c|}{$-1.22$} & $-1.10$ \\
\multicolumn{1}{|c|}{$\Xi ^0$} & $-1.25$ & $-1.40$ & \multicolumn{1}{c|}{$%
-1.33$} & $-1.37$ \\
\multicolumn{1}{|c|}{$\Xi ^{-}$} & $-0.68$ & $-0.48$ & \multicolumn{1}{c|}{$%
-0.61$} & $-0.48$ \\
\multicolumn{1}{|c|}{$\Lambda $} & $-0.61$ & $-0.59$ & \multicolumn{1}{c|}{$%
-0.59$} & $-0.60$ \\
\multicolumn{1}{|c|}{$\Lambda \Sigma $} & $-1.60$ & $-1.60$ &
\multicolumn{1}{c|}{$-1.53$} & $-1.58$ \\ \hline
\multicolumn{1}{|c|}{$\mu _u$} &  & $1.76$ & \multicolumn{1}{c|}{$2.42$} & $%
2.74$ \\
\multicolumn{1}{|c|}{$\mu _d$} &  & $-1.00$ & \multicolumn{1}{c|}{$-1.21$} &
$-1.37$ \\
\multicolumn{1}{|c|}{$\mu _s$} &  & $-0.61$ & \multicolumn{1}{c|}{$0.71$} & $%
0.82$ \\ \hline\cline{5-5}\cline{1-1}\cline{2-4}
\end{tabular}
\end{center}

\end{document}